\documentclass[12pt]{article}

\usepackage[utf8]{inputenc}
\usepackage{geometry,setspace}
\usepackage{amsmath,amssymb,amsfonts,amsbsy,bm,bbm,latexsym,mathabx}
\usepackage{natbib,hyperref}
\usepackage{graphicx,epsfig,xcolor}
\usepackage{algorithm,algpseudocode}
\usepackage{booktabs}
\usepackage{multirow}
\usepackage{changes}
\usepackage{subcaption}

\newtheorem{theorem}{Theorem}
\newtheorem{remark}{Remark}

\def\n{\nonumber}

\def\trans{^\mathrm{T}}
\def\tr{\mbox{tr}}

\def\E{\mathbb{E}}
\def\V{\mathbb{V}}
\def\sumI{\sum_{i=1}^N}
\def\wh{\widehat}
\def\wt{\widetilde}

\def\a{\mathbf{a}}
\def\x{\mathbf{x}}
\def\X{\mathbf{X}}
\def\z{\mathbf{z}}
\def\Z{\mathbf{Z}}
\def\bb{\boldsymbol{\beta}}
\def\0{\mathbf{0}}
\def\p{\mathbf{p}}
\def\A{\mathbf{A}}
\def\D{\mathbf{D}}
\def\H{\mathbf{H}}
\def\bSig{\boldsymbol{\Sigma}}

\title{Surrogate-assisted optimal sampling for risk prediction under measurement constraints}
\author{Sunhyun Park$^{1}$, Seong-ho Lee$^{1,2}$\thanks{corresponding author; email: \texttt{seongho@uos.ac.kr}}
\bigskip\\
$^1$Department of Statistics and Data Science, University of Seoul, South Korea\\
$^2$Department of Statistics, University of Seoul, South Korea}
\date{}

\begin{document}
\begin{spacing}{1.25}

\maketitle

\begin{abstract}
In many risk prediction problems, covariates and a response surrogate are routinely available for a large target population, whereas the true response is costly to ascertain and is observed only for a limited subset. 
This creates a design problem: one must decide which observations should receive response measurement in order to build a prediction model under a fixed measurement budget. 
We propose a surrogate-assisted optimal sampling framework for risk prediction under measurement constraints. 
In the target setting, the surrogate identifies confirmed positive cases, while responses for surrogate-negative observations remain unobserved and can be selectively measured, and thus the sampling design determines how the response measurement budget is allocated.
Our framework constructs an optimal sampling design minimizing the leading term of the expected out-of-sample cross-entropy loss and incorporates the resulting design into an inverse-probability-weighted cross-entropy estimator.
The proposed design depends only on covariates, the surrogate, and a preliminary estimator, and therefore does not require responses from unlabeled observations at the design stage.
We establish consistency, asymptotic normality, and leading-order prediction optimality of the resulting estimator. 
Extensive simulation studies and two real data applications demonstrate that the proposed design improves prediction performance and exhibits robustness under surrogate misspecification and rare outcome settings.
\end{abstract}


\end{spacing}


\clearpage
\section{Introduction}

Despite the availability of large-scale datasets, obtaining responses is often difficult, time-consuming, and costly \citep{settles2009active}. In many applications, there exist measurement constraints making it infeasible to obtain responses for all observations. A prominent example arises in electronic health record data \citep{zhang2020maximum}, where diagnostic codes or clinical records are easily obtainable and can serve as surrogate indicators for identifying a subset of confirmed cases, while leaving the remaining observations unlabeled. In such settings, patients without such surrogate indicators typically require manual chart review to ascertain their true disease status. Such labeling challenges also arise in many other domains. For example, labeling speech data requires expert annotation, and large-scale classification tasks involve repetitive manual labeling of massive datasets \citep{settles2009active}. In materials science, predicting the critical temperature of superconductors requires costly experiments, and in astronomy, galaxy classification relies on human visual inspection, which becomes infeasible for massive datasets \citep{hamidieh2018data, banerji2010galaxy, reiman2019deblending}. Consequently, even though all covariates are readily available, responses are often sparsely observed, severely constraining the dataset availability for machine learning \citep{wang2017computationally}.

In this problem context, it is essential to efficiently select a subset of data for labeling under measurement constraints, such as a fixed budget. This motivates the use of subsampling methods, which perform analysis on a smaller yet informative subset of data selected according to pre-designed sampling probabilities \citep{shao2025optimal}. A naive yet frequently used approach is simple random sampling (SRS), which assigns equal sampling probabilities to all observations. However, SRS fails to exploit heterogeneity in informativeness across observations, and may lead to inefficient use of limited labeling resources. In this regard, subsampling methods for linear regression have been extensively studied in the literature \citep{drineas2006sampling, ma2014statistical, wang2019information, zhang2021optimal}. Moreover, \cite{wang2018optimal} proposed a subsampling method that optimizes the asymptotic mean squared error of the estimator of the model parameter. This optimal subsampling approach has subsequently been extended to various models, including softmax regression, generalized linear models, quantile regression, and composite quantile regression \citep{yao2019optimal, ai2021optimal, wang2021optimal, shao2022optimal, shao2025optimal}. More recently, optimal sampling strategies under measurement constraints have been proposed in the literature \citep{lee2023optimal, yu2025optimal, shen2025optimal, shao2025optimal_constraints}.

We consider the setting where measuring responses is constrained, more specifically, only a subset of positive cases is identified by surrogate indicators while the remaining observations' responses are unknown. In such a setting, the observed data consist of a small group of confirmed positive cases and a large group of unlabeled observations that may contain both positive and negative cases. The main goal is to optimally select a subset of observations from this unlabeled group for response identification so that the resulting model improves out-of-sample risk prediction. In this context, the method proposed by \cite{wang2018optimal} is not directly applicable, as it requires response information to construct sampling probabilities, whereas sampling must be performed on the unlabeled group with unobserved responses. Moreover, its applicability is restricted to a specific model class. To overcome this limitation, both \cite{shen2025optimal} and \cite{lee2023optimal} propose sampling strategies that leverage surrogate variables to guide the selection of informative samples when only partial response information is available. However, their approaches still have several limitations. In particular, \cite{shen2025optimal} is developed within the generalized linear model framework and focuses on minimizing the mean squared error (MSE) of the estimator of the model parameter, thereby emphasizing parameter estimation accuracy rather than predictive performance. Similarly, \cite{lee2023optimal} considers optimal sampling strategies based on minimizing the prediction mean squared error, which are not directly aligned with binary prediction tasks.

To address this gap, we propose a novel optimal sampling framework that directly targets binary prediction performance under measurement constraints. Specifically, our contributions are threefold. First, the proposed method is applicable to a general class of smooth parametric risk prediction models, and is therefore not restricted to logistic regression or a particular generalized linear model. Second, the proposed sampling design does not require responses from the target unlabeled observations at the design stage as long as a preliminary estimator is available from a pilot sample, making it well suited to measurement-constrained scenarios. Third, we use the expected prediction cross-entropy loss as the optimization criterion, which is directly aligned with binary risk prediction and differs from existing approaches based on estimator MSE or prediction MSE. We establish the theoretical properties of the proposed method and validate its practical effectiveness through extensive numerical studies.

The remainder of this paper is organized as follows. In Section \ref{sec:methods}, we introduce the proposed sampling-based prediction model, derive the optimal sampling strategy, and establish its asymptotic properties. Section \ref{sec:sim_experiments} presents simulation studies to evaluate the performance of the proposed method under various scenarios. Section \ref{sec:realdata} illustrates the practical utility of our approach through two real data applications. Finally, Section \ref{sec:discussion} concludes with a discussion and potential directions for future research. Technical details are provided in the Supplementary Material.


\section{Proposed methods}\label{sec:methods}

\begin{table}[t]
\centering
\begin{tabular}{ll} 
\toprule
\textbf{Symbol} & \textbf{Description} \\
\midrule
$y_i$ & Response, for example, disease status; $y_i \in \{0,1\}$ \\

$s_i$ & Surrogate of $y_i$, for example, disease diagnosis; $s_i=1 \Rightarrow y_i=1$ \\

$\mathbf{z}_i$ & Covariates, for example, disease predictive information \\

$\x_i = (s_i, \mathbf{z}_i)$ & Full covariate vector \\

$(Y^*, \X^*)$ & Future observation \\

$\delta_i$ & Sampling indicator; $\delta_i=1$ for $s_i=1$, and $\delta_i \sim \mathrm{Bernoulli}(\pi_i)$ for $s_i=0$ \\

$\pi_i=\pi(\mathbf{z}_i)$ & Sampling probability for an unlabeled observation with $s_i=0$ \\

$\pi_{\rm T}(\x_i)$ & Total sampling probability; $\pi_{\rm T}(\x_i)=s_i + (1-s_i)\pi_i$ \\

$C$ & Measurement budget; $\sum_{i:s_i=0} \pi_i \le C$ \\

$\bb$ & Model parameter \\

$\bb_0$ & True model parameter \\

$\wh{\bb}_\pi$ & Estimator of the model parameter under sampling design $\pi$ \\

$p(\x, \bb)$ & Prediction model for $\Pr(Y=1 \mid \x)$ \\
\bottomrule
\end{tabular}
\caption{Description of Symbols and Notations}
\label{tab:symbols}
\end{table}

We first summarize the main symbols and notations used throughout the paper in Table \ref{tab:symbols}. To illustrate our setting, we provide an example based on electronic health records (EHR). $y_i$ corresponds to a binary indicator of disease status, where $y_i=1$ if the $i$th patient has the disease and $y_i=0$ otherwise. $s_i$ is a surrogate of $y_i$, such that $s_i=1$ implies $y_i=1$. Suppose that the research goal is to predict the presence of depression, then $y_i$ indicates whether the $i$th patient truly has depression, and $s_i$ would represent whether the patient has been diagnosed with depression. In addition, $\mathbf{z}_i$ denotes additional covariates that provide predictive information for the disease risk.

\subsection{Sampling-based risk prediction model}

Our goal is to construct a prediction model $p(\x, \bb)$ for the probability $\Pr(Y=1 \mid \x)$ based on the covariate vector $\mathbf{x}_i = (s_i, \mathbf{z}_i)$. At the initial stage, all $\mathbf{x}_i$ $(i=1,\dots,N)$ are available from the data, whereas the true responses $y_i$ would not be fully observed, since not all observations have been verified for the response. Therefore, prior to building the prediction model, it is necessary to selectively identify $y_i$ for a subset of observations, after which the model can be trained using the observed pairs $(y_i, \mathbf{x}_i)$.

Since $s_i=1$ implies $y_i=1$, the responses of observations with $s_i=1$ are readily identified. However, these observations alone are insufficient for training a binary prediction model because they contain no confirmed negative cases. Therefore, it is necessary to sample observations with $s_i=0$ and identify their responses so that both positive and negative cases are represented in the training data. For this purpose, let $\pi(\z_i)$ be the probability of sampling the $i$th observation with $s_i=0$ and $\delta_i$ be the indicator of whether the $i$th observation is chosen for response identification. In summary, the observed data can be coded as $(\delta_i,\delta_iy_i,s_i,\z_i)~(i=1,\dots,N)$.

Once both the positive ($y_i=1$) and negative ($y_i=0$) cases are collected, we can build the prediction model $p(\cdot,\bb)$ for $y$. To this end, we propose to estimate the prediction model parameter $\bb$ via optimizing the inverse probability weighted cross-entropy loss
\begin{align}\label{eq:loss}
    \mathcal{L}_N(\bb)=-\sumI\frac{\delta_i}{\pi_{\rm T}(\x_i)}\left[y_i\log p(\x_i,\bb)+(1-y_i)\log\{1-p(\x_i,\bb)\}\right],
\end{align}
where $\pi_{\rm T}(\x_i)$ represents the probability of sampling the $i$th observation, defined as
\begin{align*}
    \pi_{\rm T}(\x_i)=s_i+(1-s_i)\pi(\z_i).
\end{align*}
We define $\pi_{\rm T}(\cdot)$ as above because the response $y_i$ of observations with $s_i=1$ is already known to be $y_i=1$ thus $\pi_{\rm T}(\x_i)=1$, while $y_i$ of observations with $s_i=0$ is identified with probability $\pi(\z_i)$. The inverse-probability weighting is used because $\E\{\delta_i/\pi_{\rm T}(\x_i)\mid y_i,\x_i\}=1$, so the weighted loss targets the full-data cross-entropy loss while accounting for the sampling design. Let us denote the optimizer of \eqref{eq:loss} as $\wh\bb_\pi$, where the subscript $\pi$ indicates its dependency on how the sampling probability $\pi(\cdot)$ is designed. The entire algorithm is summarized in Algorithm \ref{alg:sampling}.

\begin{algorithm}[t]
\caption{Sampling-based risk prediction model}\label{alg:sampling}
\begin{algorithmic}
  \State \textbf{Input data}: $\{\x_i=(s_i ,\z_i):i=1,\dots,N\}$.
  \State \textbf{Input model}: A sampling probability $\pi(\cdot)$, a risk prediction model $p(\cdot,\bb)$.
  \State \textbf{Output}: An estimated model $p(\cdot,\wh\bb_\pi)$.
  \State \textbf{Do}:
  \State (a) For $i=1,\dots,N$;\\
  \phantom{(a)} if $s_i=1$, assign $\delta_i\leftarrow 1$ and $y_i\leftarrow 1$;\\
  \phantom{(a)} if $s_i=0$, draw $\delta_i\sim\mbox{Ber}\{\pi(\z_i)\}$, and if $\delta_i=1$, identify $y_i$.
  \State (b) Construct the observed data $\{(\delta_i,\delta_iy_i,s_i ,\z_i):i=1,\dots,N\}$.
  \State (c) Estimate $\wh\bb_\pi$ via optimizing $\mathcal{L}_N(\bb)$ in \eqref{eq:loss}.
\end{algorithmic}
\end{algorithm}

We demonstrate the asymptotic properties of the estimator of the model parameter, $\wh\bb_\pi$. Let us denote the true model parameter as $\bb_0$, and the derivative of $p(\x,\bb)$ with respect to $\bb$ as $\p_{\bb}'(\x,\bb)$. Detailed proofs are provided in the Supplementary Material.

\begin{enumerate}
\renewcommand{\labelenumi}{\theenumi}
\renewcommand{\theenumi}{(C\arabic{enumi})}
    \item\label{con:bb0}
    $\bb_0 \in \operatorname{int}(\Theta)$, where $\Theta$ is compact, and $\epsilon_p<p(\x,\bb_0)<1-\epsilon_p$ for some constant $\epsilon_p>0$.
    \item\label{con:p2}
    $p(\mathbf{x}, \bb)$ is twice continuously differentiable with respect to $\bb \in \Theta$. $p(\mathbf{x}, \bb)$, $\mathbf{p}'_{\bb}(\mathbf{x}, \bb)$, and $\mathbf{p}''_{\bb\bb\trans}(\mathbf{x}, \bb)$ are equicontinuous with respect to $\bb \in \Theta$, and satisfy $\|\mathbf{p}'_{\bb}(\mathbf{x}, \bb)\|_2 < \infty$ and $\|\mathbf{p}''_{\bb\bb\trans}(\mathbf{x}, \bb)\|_2 < \infty$. In addition, the $\ell_2$-norm of the 3rd order derivative of $p(\x,\bb)$ with respect to $\bb$ is bounded.
    \item\label{con:grad}
    The population cross-entropy loss has a unique minimizer at $\bb_0$. In addition, $\D=\E\left({\p_{\bb}'}^{\otimes2}(\X,\bb_0)/[p(\X,\bb_0)\{1-p(\X,\bb_0)\}]\right)$ is invertible, and its minimum eigenvalue is greater than $\epsilon_\D$, where $\epsilon_\D$ is a positive constant.
    \item\label{con:pi}
    The sampling probabilities are admissible in the sense that $\epsilon_\pi \le \pi(\z_i)\le 1$ for some constant $\epsilon_\pi>0$.
\end{enumerate}

\begin{theorem}\label{th:consistency}
    Under Conditions \ref{con:bb0}--\ref{con:pi}, $\wh\bb_\pi$ converges in probability to $\bb_0$ as $N\to\infty$.
\end{theorem}

\begin{theorem}\label{th:distribution}
    Under Conditions \ref{con:bb0}--\ref{con:pi}, $N^{1/2}(\wh\bb_\pi-\bb_0)\to N(\0,\D^{-1}\bSig_\pi\D^{-1})$ in distribution as $N\to\infty$, where
    \begin{align*}
        \bSig_\pi=\E\left[\frac{1}{\pi_{\rm T}(\X)}\frac{{\p_{\bb}'}^{\otimes2}(\X,\bb_0)}{p(\X,\bb_0)\{1-p(\X,\bb_0)\}}\right].
    \end{align*}
\end{theorem}


\subsection{Prediction-optimal sampling}

As demonstrated in Theorem \ref{th:distribution}, the asymptotic behavior of $\wh\bb_\pi$ depends on the sampling probability $\pi(\cdot)$, and so does the prediction performance of $p(\cdot,\wh\bb_\pi)$. Therefore, if we can identify the leading contribution of $\pi(\cdot)$ to the prediction risk, we can design the sampling probability to improve the resulting out-of-sample prediction performance.

For a future observation $(Y^*,\X^*)$, we define the expected cross-entropy loss of a subsampling-based prediction model $p(\cdot,\wh\bb_\pi)$ as
\begin{align*}
    Q(\wh\bb_\pi)
    =
    -\E\left[Y^*\log p(\X^*,\wh\bb_\pi)+(1-Y^*)\log\{1-p(\X^*,\wh\bb_\pi)\}\right].
\end{align*}
A second-order expansion around $\bb_0$, together with Theorem \ref{th:distribution}, shows that the leading $N^{-1}$ term of $Q(\wh\bb_\pi)-Q(\bb_0)$ depends on $\pi(\cdot)$ through

\begin{align}\label{eq:target}
    \E\left[\frac{1}{\pi(\Z)}\frac{{\p_{\bb}'}\trans(\X,\bb_0)\D^{-1}{\p_{\bb}'}(\X,\bb_0)}{p(\X,\bb_0)\{1-p(\X,\bb_0)\}}\mid S=0\right].
\end{align}

Hence, optimizing this term with respect to $\pi(\cdot)$ leads to optimizing the leading contribution to the expected cross-entropy loss of the estimated prediction model $p(\cdot,\wh\bb_\pi)$. Since Equation \eqref{eq:target} is theoretical and therefore unknown in practice, we propose an empirical approximation and choose the sampling probabilities $\wh\pi_i$ by minimizing the empirical counterpart of \eqref{eq:target} under the measurement constraint.

Let
\begin{align*}
    a_i(\wt\bb)
    =
    \frac{{\p_{\bb}'}\trans(\x_i,\wt\bb)\wt\D^{-1}{\p_{\bb}'}(\x_i,\wt\bb)}
    {p(\x_i,\wt\bb)\{1-p(\x_i,\wt\bb)\}},
    \qquad i:s_i=0.
\end{align*}
We choose
\begin{align}\label{eq:optimalprob}
    \wh{\pi}
    =
    {\arg\min}_{\{\pi_i: s_i=0\}}
    \sum_{i:s_i=0}\frac{a_i(\wt\bb)}{\pi_i}
    \quad\text{subject to}\quad
    \sum_{i:s_i=0}\pi_i\leq C,
    \quad
    \epsilon_\pi\leq \pi_i\leq 1.
\end{align}
When the upper and lower bounds are inactive, the solution has the simple form
\begin{align*}
    \wh\pi_i
    =
    C\frac{\sqrt{a_i(\wt\bb)}}{\sum_{j:s_j=0}\sqrt{a_j(\wt\bb)}} ,
    \qquad i:s_i=0,
\end{align*}

where $\wt\bb$ is an estimate of $\bb_0$ and $\wt\D=N^{-1}\sumI{\p_{\bb}'}^{\otimes2}(\x_i,\wt\bb)/[p(\x_i,\wt\bb)\{1-p(\x_i,\wt\bb)\}]$. The algorithm is summarized in Algorithm \ref{alg:optimal}.

\begin{algorithm}[t]
\caption{Optimal sampling-based risk prediction model}\label{alg:optimal}
\begin{algorithmic}
  \State \textbf{Input data}: $\{\x_i=(s_i ,\z_i):i=1,\dots,N\}$, a measurement constraint $C$, an estimate $\wt\bb$.
  \State \textbf{Input model}: a risk prediction model $p(\cdot,\bb)$.
  \State \textbf{Output}: An estimated model $p(\cdot,\wh\bb_{\wh{\pi}})$.
  \State \textbf{Do}:
  \State (a) Obtain $\wh{\pi}$ as \eqref{eq:optimalprob}.
  \State (b) For $i=1,\dots,N$, if $s_i=0$, assign $\pi(\z_i)\leftarrow\wh\pi_i$.
  \State (c) Implement Algorithm \ref{alg:sampling}.
\end{algorithmic}
\end{algorithm}

We make a note on the choice of $\wt\bb$ for the initialization of Algorithm \ref{alg:optimal}. To ensure that the optimal sampling probability $\wh{\pi}$ optimizes the prediction cross-entropy loss of the estimated model $p(\cdot,\wh\bb_{\pi})$ asymptotically, $\wt\bb$ has to be consistent for $\bb_0$. For the consistency of $\wt\bb$, one may adopt a result from a pilot study, or use a data set collected in advance to the current study. If no prior information is available, we may first implement Algorithm \ref{alg:sampling} with equal probability $\pi(\z_i)=\pi_0$ to establish an initial estimate $\wt\bb=\wh\bb_{\pi_0}$ which is consistent for $\bb_0$ according to Theorem \ref{th:consistency}, then utilize this estimate to further implement Algorithm \ref{alg:optimal}. This procedure is summarized in Algorithm \ref{alg:twostep}.

\begin{algorithm}[t]
\caption{Two-step procedure}\label{alg:twostep}
\begin{algorithmic}
  \State \textbf{Input data}: $\{\x_i=(s_i ,\z_i):i=1,\dots,N\}$, a measurement constraint $C$.
  \State \textbf{Input model}: a constant sampling probability $\pi_0$, a risk prediction model $p(\cdot,\bb)$.
  \State \textbf{Output}: An estimated model $p(\cdot,\wh\bb_{\wh{\pi}})$.
  \State \textbf{Do}:
  \State (a) Assign $\pi(\cdot)\leftarrow\pi_0$.
  \State (b) Implement Algorithm \ref{alg:sampling} and assign $\wt\bb\leftarrow\wh\bb_{\pi_0}$.
  \State (c) Implement Algorithm \ref{alg:optimal}.
\end{algorithmic}
\end{algorithm}

We make a remark on how our proposed method compares to existing methods. Our method is designed for optimizing the prediction cross-entropy of a smooth parametric prediction model. On the other hand, the method by \cite{wang2018optimal} optimizes the asymptotic variance of the model parameter estimator under a logistic regression model. Thus, our method is preferable if one is interested in risk prediction, and it is not restricted to logistic regression. In addition, \cite{wang2018optimal} requires that $y_i$ is observed in advance to design the sampling probability, while designing our optimal sampling probability does not involve $y_i$, hence is more applicable under the problem context. Although the method by \cite{lee2023optimal} also enjoys these advantages, their method optimizes the mean squared error rather than the cross-entropy, thus our choice is more directly aligned with binary prediction. We will further demonstrate the advantage of our method over \cite{lee2023optimal} through numerical examples in the following sections.

Now, we demonstrate the optimality of prediction model $p(\cdot,\wh\bb_{\wh{\pi}})$ among all sampling-based prediction models $p(\cdot,\wh\bb_\pi)$ under the measurement constraint in terms of the expected cross-entropy loss. A detailed proof is provided in the Supplementary Material.

\begin{theorem}\label{th:optimality}
Assume Conditions \ref{con:bb0}--\ref{con:pi}. 
Let $\wh\bb_\pi$ be the estimator obtained by minimizing Equation \eqref{eq:loss} under any admissible sampling strategy $\pi$ satisfying the measurement constraint $\sum_{i:s_i=0}\pi_i \le C$. 
Let $\wh{\bb}_{\wh{\pi}}$ be the estimator obtained under the optimal sampling probabilities $\wh{\pi}$ defined in Equation \eqref{eq:optimalprob}. Then,
\begin{align*}
Q(\wh\bb_{\wh\pi})
\le
Q(\wh\bb_{\pi})+o_p^+(N^{-1}),
\end{align*}
where
\begin{align*}
Q(\bb)=
-\E\big[Y^*\log p(\X^*,\bb)+(1-Y^*)\log\{1-p(\X^*,\bb)\}\big].
\end{align*}
\end{theorem}

\begin{remark}
Theorem \ref{th:optimality} links the empirical sampling design in Equation \eqref{eq:optimalprob} to the target prediction criterion $Q(\cdot)$. In particular, among admissible sampling designs with the same measurement budget, the proposed design minimizes the leading contribution to the expected out-of-sample cross-entropy loss. This supports the usefulness of the cross-entropy-based sampling criterion when the primary goal is binary risk prediction rather than parameter estimation or mean-squared prediction error.
\end{remark}


\section{Simulation experiments} \label{sec:sim_experiments}

We conduct simulation studies to assess the finite-sample performance of our proposed method. For comparison, we implement five estimation procedures: the proposed cross-entropy based optimal sampling (CE), the mean-squared-error based optimal sampling of \cite{lee2023optimal} (MSE), the estimation procedure proposed by \cite{yin2021cost} (OSCA), the maximum likelihood estimator under simple random sampling (SRS), as well as TRUE that serves as a benchmark evaluated under the true underlying parameter.

All results are based on $1,000$ independent replicates with a total sample size of $N=5,000$ for each replicate. In addition, we consider different resource constraints by setting $C=200, 300$, and $400$ to compare the performances under varying limits of available sampling budgets. For the initial estimator $\wt{\bb}$, we use 400 observations drawn from the same population and independent of the $N$ observations. Overall, the average response prevalence is fixed at $\mathbb{E}\{p_0(\mathbf{X})\}=0.2$, and the surrogate sensitivity is set to approximately $40\%$, that is, $\Pr(S=1 \mid Y=1)=0.4$. In the following, the detailed settings were borrowed from \cite{lee2023optimal}.

We primarily evaluate model performance using the cross-entropy loss (CE), which directly reflects the optimization objective of the proposed estimator:
$$
\mathrm{CE} = -\sum_{i=1}^{N} \bigl[y_i \log p(\mathbf{x}_i, \wh{\bb}) + (1 - y_i)\log\{1 - p(\mathbf{x}_i, \wh{\bb})\}\bigr].
$$
For numerical stability, the reported CE is computed using observations whose predicted probabilities lie within $[10^{-10}, 1-10^{-10}]$, and then rescaled by $N/N_{\epsilon}$, where $N_{\epsilon}$ denotes the number of such observations. For comparison, we also report the mean squared error (MSE), specificity (TN), sensitivity (TP), and the area under the receiver operating characteristic curve (AUC). Specificity and sensitivity are computed using the decision rule $I\{p(\mathbf{x}, \wh{\bb}) > 0.5\}$.


\subsection{When surrogates correctly detect positive cases}\label{sec:sim1}

We first consider the situation where the surrogates correctly detect the positive cases since $\Pr(Y=1 \mid S=1)=1$. The covariate vector $\mathbf{Z}$ is of dimension $12$, where $Z_1, Z_5, Z_9$ are independently generated from $N(0,1)$; $Z_2, Z_6, Z_{10}$ are generated from a discrete uniform distribution on $\{1, 2, 3, 4, 5\}$; $Z_3, Z_7, Z_{11}$ follow a Bernoulli distribution with success probability $p=0.5$; and $Z_4, Z_8, Z_{12}$ follow a Bernoulli distribution with $p=0.1$, representing categorical variables with low prevalence. The regression coefficient is set to $\bb = (0.2, 0.3, 0.4, -0.5, 0.8, 1.0, -1.2, 1.4, 1.7, -2.0, 2.3, 2.6)^\top$, so that $Z_1, Z_2, Z_3, Z_4$ represent weak, $Z_5, Z_6, Z_7, Z_8$ moderate, and $Z_9, Z_{10}, Z_{11}, Z_{12}$ strong covariates.

The response $Y_i$ is generated from a logistic regression model $p_0(\mathbf{x}_i) = p(\mathbf{z}_i, \bb) = \mathrm{expit}(\beta_0 + \bb^\top \mathbf{z}_i),$ where $\mathrm{expit}(\eta) = 1/(1+e^{-\eta})$, and $\beta_0$ is determined to yield response prevalence of $20\%$. The surrogate $S_i$ is generated as $S_i=Y_i B_i$, where $B_i\mid \mathbf{Z}_i\sim \mathrm{Bernoulli}(p_i)$ with $p_i = \text{expit}(\gamma_0 + \boldsymbol{\gamma}^{\mathsf T} \mathbf{z}_i) / \text{expit}(\beta_0 + \bb^{\mathsf T} \mathbf{z}_i)$. Here, $\boldsymbol{\gamma} = \bb + 0.1\, \mathrm{sign}(\bb)$, and $\gamma_0$ is chosen so that $\Pr(S=1\mid Y=1)\approx 0.4$.

\begin{table}[t]
\centering
\caption{Summary of simulation experiment in Section \ref{sec:sim1}}
\label{tab:combined_results}
\begin{tabular}{llrrrrr}
\toprule
\textbf{C} & \textbf{Method} & \textbf{CE} & \textbf{MSE} & \textbf{TN} & \textbf{TP} & \textbf{AUC} \\
\midrule
\multirow{5}{*}{200}
& CE   & 1144.044 & 0.069 & 0.948 & 0.725 & 0.951 \\
& MSE  & 8372.493 & 0.088 & 0.940 & 0.723 & 0.943 \\
& OSCA & 1662.138 & 0.100 & 0.863 & 0.857 & 0.936 \\
& SRS  & 1445.803 & 0.087 & 0.880 & 0.881 & 0.954 \\
& TRUE & 1022.575 & 0.063 & 0.957 & 0.732 & 0.958 \\
\midrule

\multirow{5}{*}{300}
& CE   & 1104.451 & 0.068 & 0.950 & 0.728 & 0.953 \\
& MSE  & 3758.192 & 0.076 & 0.947 & 0.726 & 0.949 \\
& OSCA & 1314.562 & 0.082 & 0.896 & 0.843 & 0.950 \\
& SRS  & 1291.369 & 0.079 & 0.898 & 0.862 & 0.955 \\
& TRUE & 1022.575 & 0.063 & 0.957 & 0.732 & 0.958 \\
\midrule

\multirow{5}{*}{400}
& CE   & 1079.864 & 0.066 & 0.952 & 0.728 & 0.954 \\
& MSE  & 1704.559 & 0.070 & 0.950 & 0.730 & 0.952 \\
& OSCA & 1190.462 & 0.074 & 0.913 & 0.827 & 0.953 \\
& SRS  & 1211.768 & 0.074 & 0.909 & 0.847 & 0.956 \\
& TRUE & 1022.575 & 0.063 & 0.957 & 0.732 & 0.958 \\
\bottomrule
\end{tabular}
\end{table}

In Table \ref{tab:combined_results}, we report the performance of the implemented methods under different sampling budgets ($C=200, 300, 400$). We observe that the proposed $\text{CE}$ method achieves the lowest cross-entropy loss ($\text{CE}$) across all conditions, clearly outperforming $\text{MSE}$, $\text{OSCA}$, and $\text{SRS}$ in its primary evaluation metric. Interestingly, $\text{CE}$ also attains the smallest $\text{MSE}$ values, suggesting that optimizing the cross-entropy criterion indirectly improves mean squared error performance as well. In terms of classification characteristics, $\text{CE}$ maintains a well-balanced trade-off between specificity ($\text{TN}$) and sensitivity ($\text{TP}$), close to the performance of the $\text{TRUE}$ model. The performance differences among methods become more pronounced as the sampling budget decreases, indicating that $\text{CE}$ performs more reliably under resource-constrained conditions. Overall, $\text{CE}$ provides stable and efficient estimation across all settings, with particularly notable improvements when the sampling budget is limited.


\subsection{When surrogates incorrectly alarm positive cases}\label{sec:sim2}

Now, to assess the sensitivity of the proposed method to the positive-only surrogate assumption, we consider a scenario in which the true data-generating structure does not satisfy $\Pr(Y=1 \mid S=1)=1$. The covariates $\mathbf{Z}_i$ and the response $Y_i$ are generated in the same manner as described in Section \ref{sec:sim1}. The surrogate $S_i$ is redefined as $S_i = Y_i S_{1i} + (1 - Y_i) S_{0i},$ where $S_{1i} \sim \text{Bernoulli}(p_i)$ and $S_{0i} \sim \text{Bernoulli}(q)$. Here, $p_i = \text{expit}(\gamma_0 + \boldsymbol{\gamma}^{\mathsf T} \mathbf{z}_i) / \text{expit}(\beta_0 + \bb^{\mathsf T} \mathbf{z}_i)$, $q = 0.025$, and $\boldsymbol{\gamma} = \bb + 0.1\,\text{sign}(\bb)$. The intercepts $\beta_0$ and $\gamma_0$ are determined such that the response prevalence is $20\%$ and the surrogate sensitivity is approximately $40\%$. With this choice, approximately $20\%$ of surrogate-positive observations are false positives, that is, $\Pr(Y=1 \mid S=1)\approx 0.8$. All estimation procedures are implemented under the false assumption that $\Pr(Y=1 \mid S=1)=1$, meaning that all observations with $S=1$ are incorrectly labeled as positive ($Y=1$). Each method is then fitted using a logistic regression model of $Y$ on $\mathbf{Z}$, and the prediction performance is evaluated with respect to the true response $Y$.

\begin{table}[t]
\centering
\caption{Summary of simulation experiment in Section \ref{sec:sim2}}
\label{tab:combined_results_robustness}
\begin{tabular}{llrrrrr}
\toprule
\textbf{C} & \textbf{Method} & \textbf{CE} & \textbf{MSE} & \textbf{TN} & \textbf{TP} & \textbf{AUC} \\
\midrule
\multirow{5}{*}{200}
& CE   & 1144.968 & 0.070 & 0.946 & 0.730 & 0.951 \\
& MSE  & 6477.654 & 0.085 & 0.937 & 0.735 & 0.944 \\
& OSCA & 3153.756 & 0.222 & 0.536 & 0.989 & 0.934 \\
& SRS  & 3407.929 & 0.242 & 0.510 & 0.994 & 0.943 \\
& TRUE & 1022.575 & 0.063 & 0.957 & 0.732 & 0.958 \\
\midrule

\multirow{5}{*}{300}
& CE   & 1128.553 & 0.069 & 0.948 & 0.733 & 0.952 \\
& MSE  & 2556.290 & 0.074 & 0.944 & 0.738 & 0.949 \\
& OSCA & 2490.668 & 0.166 & 0.688 & 0.972 & 0.945 \\
& SRS  & 2677.361 & 0.183 & 0.641 & 0.984 & 0.948 \\
& TRUE & 1022.575 & 0.063 & 0.957 & 0.732 & 0.958 \\
\midrule

\multirow{5}{*}{400}
& CE   & 1121.401 & 0.068 & 0.949 & 0.735 & 0.953 \\
& MSE  & 1456.396 & 0.069 & 0.947 & 0.738 & 0.952 \\
& OSCA & 2136.697 & 0.136 & 0.769 & 0.951 & 0.949 \\
& SRS  & 2273.057 & 0.150 & 0.720 & 0.970 & 0.950 \\
& TRUE & 1022.575 & 0.063 & 0.957 & 0.732 & 0.958 \\
\bottomrule
\end{tabular}
\end{table}

Table \ref{tab:combined_results_robustness} summarizes the performance of each implemented method when the positive-only assumption is violated. The proposed $\text{CE}$ method achieves the lowest cross-entropy loss across all sampling budgets ($C=200, 300, 400$), demonstrating superior performance compared to $\text{MSE}$, $\text{OSCA}$, and $\text{SRS}$ even when the surrogate assumption is violated. The $\text{CE}$ method also exhibits the smallest MSE, suggesting that optimizing the cross-entropy criterion can lead to favorable mean squared error performance in this setting. In contrast to Table \ref{tab:combined_results}, where the $\text{SRS}$ method recorded the highest $\text{AUC}$, this advantage disappears in this mislabeling scenario. In contrast, the CE and MSE methods appear more robust to false-positive contamination, partly because the inverse-probability-weighted objective gives relatively larger influence to verified observations sampled from the $S=0$ group than to surrogate-labeled positive observations. The performance gap further widens as the sampling budget decreases, confirming that the $\text{CE}$ method operates reliably even in resource-constrained environments. Overall, the $\text{CE}$ method maintains empirical robustness when the positive-only assumption is violated, preserving both loss minimization efficiency and classification stability despite the presence of a false-positive surrogate.


\section{Real data application}\label{sec:realdata}

To evaluate the practical performance of the proposed method in real-world environments, we applied the proposed method (CE) to two datasets. For comparison, we included the MSE-based method of \cite{lee2023optimal}, the OSCA method of \cite{yin2021cost}, the maximum likelihood estimator under simple random sampling (SRS), and the FULL estimator obtained by minimizing cross-entropy using the entire dataset. 
Across both datasets, the same evaluation metrics were used: cross-entropy loss (CE), mean squared error (MSE), specificity (TN), sensitivity (TP), and the area under the ROC curve (AUC). All metrics were computed excluding the initial $m$ observations used for pilot estimation, consistent with the simulation setting described in Section \ref{sec:sim_experiments}.

\subsection{Depression prediction}\label{sec:data_example1}

The data were obtained from the Medical Information Mart for Intensive Care III database version 1.4 (MIMIC-III v1.4) by \cite{johnson2016mimic}, a publicly available database containing clinical information on more than 40,000 patients admitted to the intensive care units of the Beth Israel Deaconess Medical Center between 2001 and 2012. Among these, we extracted 767 admission records corresponding to 415 patients with repeated ICU admissions. The same set of patients and admission records was also analyzed in \cite{gehrmann2018comparing}.

The objective of this application is to classify each admission record into whether the patient exhibited depression symptoms. Accordingly, we defined the label $Y=1$ if depression symptoms were present. We further assigned the surrogate $S=1$ if depression-related diagnosis was recorded based on ICD-9 diagnostic codes. According to the cross-validated chart reviews by clinical experts in \cite{gehrmann2018comparing}, 30.7\% of the admission records were confirmed to exhibit depression symptoms ($Y=1$), and among these, only 24.4\% were surrogate-positive ($S=1$), indicating the low sensitivity of ICD-based depression diagnosis codes. In addition, 5 observations with $S=1$ but $Y=0$ were identified, violating the positive-only assumption; therefore, these were removed from the analysis. This analysis setting was borrowed from \cite{lee2023optimal}.
 
For the analysis, we adopted a logistic regression model with covariates including gender, age and its quadratic term, ethnicity, marital status, insurance type, admission type, previous length of stay, the number of procedures in each category, and the number of prescribed antidepressant medications. The initial estimator $\wt{\bb}$ was obtained by randomly sampling $m=100$ observations from the full dataset, after which each method was applied to the remaining records. The surrogate-positive probability $\Pr(S=1 \mid \Z)$ was also estimated via logistic regression using the same covariates. A total of 1,000 replicates were conducted under sampling budgets $C = 100, 200, 300, 400$.

\begin{figure}[t]
    \centering
    \includegraphics[width=0.7\textwidth]{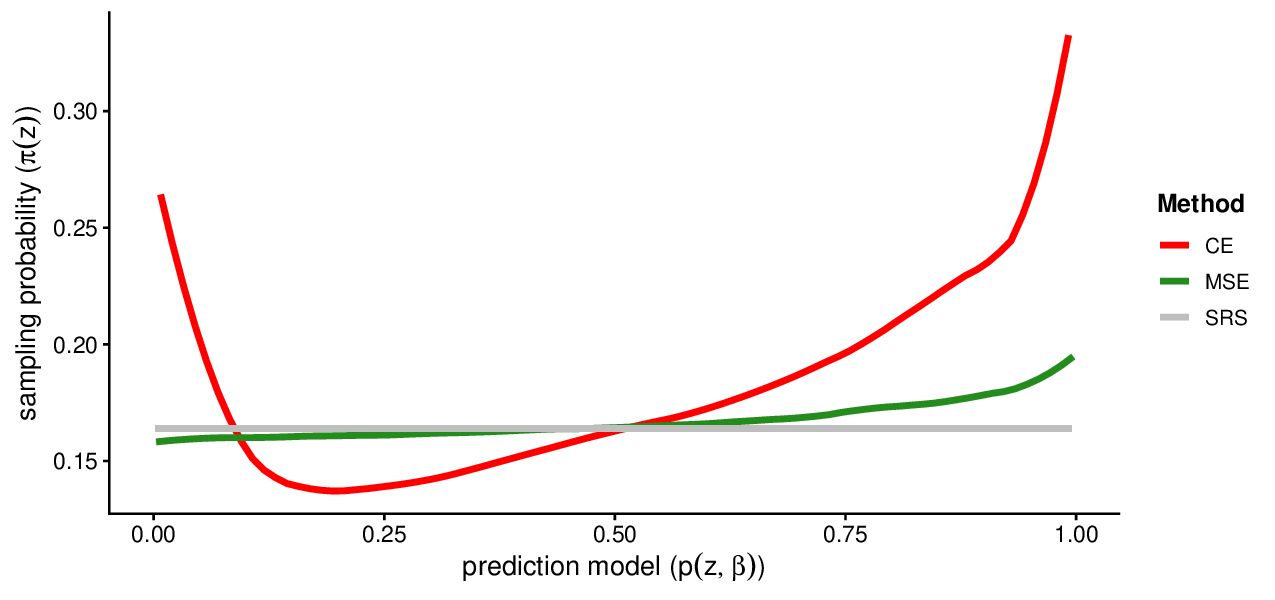}
    \caption{Sampling probability $\pi(\z)$ under $C=100$ from the real data application in Section \ref{sec:data_example1}.}
    \label{fig:sampling100}
\end{figure}

\begin{figure}[t]
    \centering
    \begin{subfigure}{0.32\textwidth}
        \includegraphics[width=\textwidth]{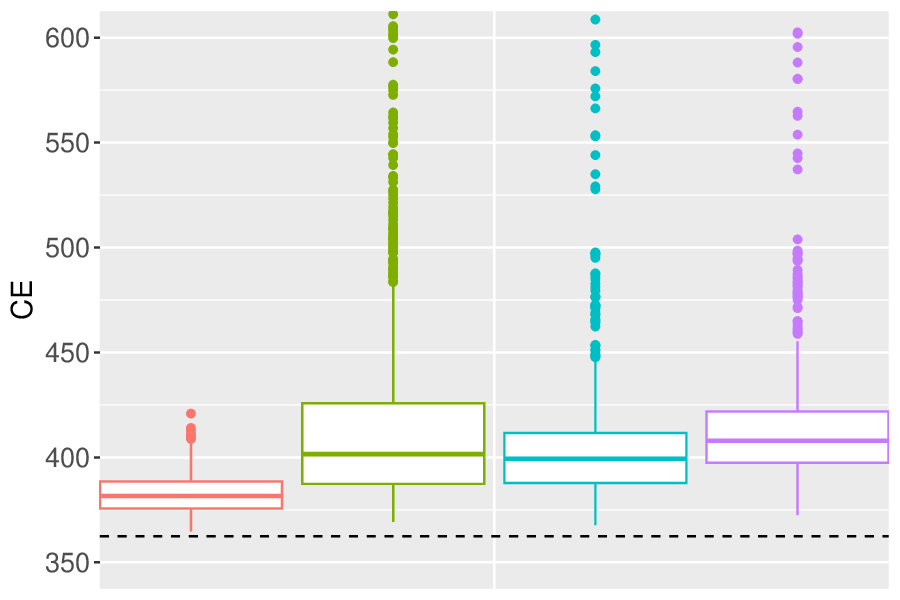}
        \caption{$C=200$}
        \end{subfigure}
        \hfill
        \begin{subfigure}{0.32\textwidth}
        \includegraphics[width=\textwidth]{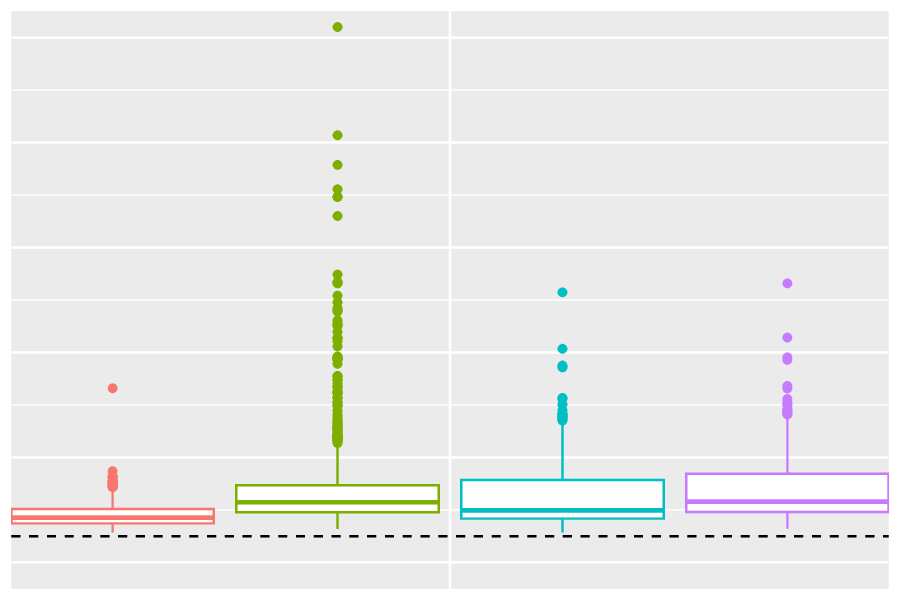}
        \caption{$C=300$}
        \end{subfigure}
        \hfill
        \begin{subfigure}{0.32\textwidth}
        \includegraphics[width=\textwidth]{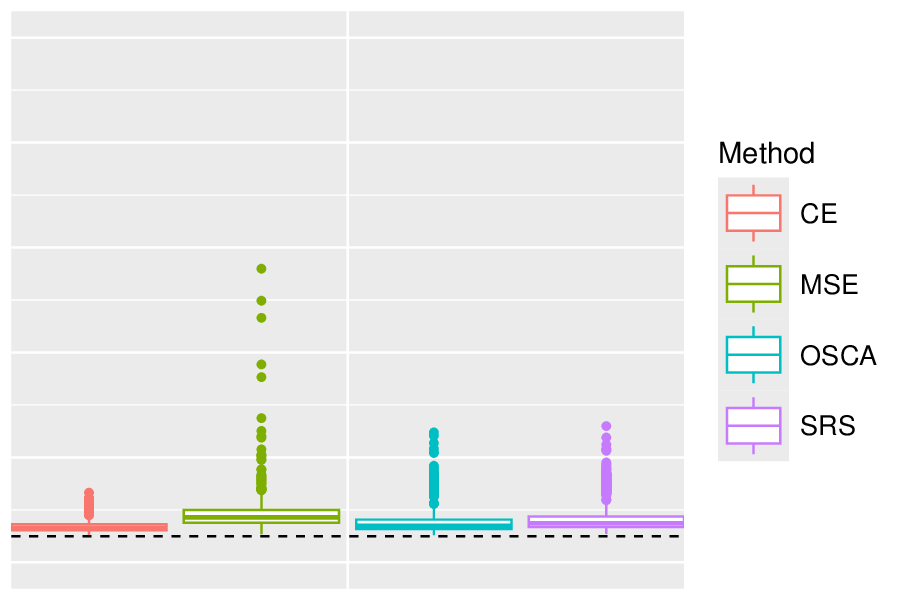}
        \caption{$C=400$}
    \end{subfigure}
    \caption{Boxplots of the cross-entropy loss (CE) from the real data application in Section \ref{sec:data_example1}. Dashed: CE based on the full dataset.}
    \label{fig:CE_plot}
\end{figure}

Figure~\ref{fig:sampling100} shows how sampling probabilities $\pi(\z)$ vary with the predicted probability $p(\z, \bb)$ among observations with $S=0$ across different methods, using LOESS-smoothed curves obtained from 1,000 simulation replicates. The proposed CE method assigns larger probabilities to observations with larger estimated contributions to the leading cross-entropy risk, as quantified by $a_i(\wt\bb)$ in Equation \eqref{eq:optimalprob}, whereas SRS assigns constant probabilities. These results illustrate that the proposed method prioritizes observations that are more informative for the target prediction loss rather than treating all unlabeled observations equally.

Figure~\ref{fig:CE_plot} presents the boxplots of the cross-entropy loss across different sampling methods under varying sampling budgets. The proposed CE method consistently achieves lower median and variability compared to other methods, demonstrating its efficiency across different budget levels.

\begin{table}[t]
\centering
\caption{Summary of real data application in Section \ref{sec:data_example1}}
\label{tab:combined_results_realdata1}
\begin{tabular}{llrrrrr}
\toprule
\textbf{C} & \textbf{Method} & \textbf{CE} & \textbf{MSE} & \textbf{TN} & \textbf{TP} & \textbf{AUC} \\
\midrule
\multirow{5}{*}{100}
& CE   & 416.815 & 0.203 & 0.872 & 0.352 & 0.678 \\
& MSE  & 895.806 & 0.236 & 0.838 & 0.421 & 0.655 \\
& OSCA & 491.558 & 0.217 & 0.767 & 0.514 & 0.679 \\
& SRS  & 517.179 & 0.230 & 0.710 & 0.572 & 0.679 \\
& FULL & 362.451 & 0.182 & 0.924 & 0.299 & 0.728 \\
\midrule

\multirow{5}{*}{200}
& CE   & 382.833 & 0.190 & 0.909 & 0.324 & 0.701 \\
& MSE  & 422.655 & 0.195 & 0.892 & 0.369 & 0.693 \\
& OSCA & 404.724 & 0.192 & 0.878 & 0.394 & 0.700 \\
& SRS  & 413.932 & 0.197 & 0.838 & 0.459 & 0.700 \\
& FULL & 362.451 & 0.182 & 0.924 & 0.299 & 0.728 \\
\midrule

\multirow{5}{*}{300}
& CE   & 372.485 & 0.186 & 0.920 & 0.312 & 0.712 \\
& MSE  & 385.077 & 0.187 & 0.909 & 0.350 & 0.710 \\
& OSCA & 380.094 & 0.186 & 0.910 & 0.343 & 0.711 \\
& SRS  & 383.487 & 0.188 & 0.886 & 0.400 & 0.710 \\
& FULL & 362.451 & 0.182 & 0.924 & 0.299 & 0.728 \\
\midrule

\multirow{5}{*}{400}
& CE   & 367.270 & 0.184 & 0.925 & 0.306 & 0.719 \\
& MSE  & 373.058 & 0.183 & 0.915 & 0.341 & 0.718 \\
& OSCA & 370.309 & 0.184 & 0.921 & 0.321 & 0.718 \\
& SRS  & 371.426 & 0.184 & 0.907 & 0.361 & 0.717 \\
& FULL & 362.451 & 0.182 & 0.924 & 0.299 & 0.728 \\
\bottomrule
\end{tabular}
\end{table}

In Table \ref{tab:combined_results_realdata1}, we summarize the performance of the proposed and competing methods under different sampling budgets ($C = 100, 200, 300, 400$). Across all settings, the proposed $\text{CE}$ method consistently achieves the lowest cross-entropy loss, outperforming $\text{MSE}$, $\text{OSCA}$, and $\text{SRS}$ in its primary optimization criterion. The MSE values of $\text{CE}$ are also close to the best competing method across all budgets, although the main advantage of the proposed method is most clearly reflected in the cross-entropy loss. In terms of classification characteristics, $\text{CE}$ delivers competitive and stable performance in specificity ($\text{TN}$), sensitivity ($\text{TP}$), and discrimination ($\text{AUC}$). Notably, when the sampling budget reaches $C \ge 200$, $\text{CE}$ achieves the highest $\text{AUC}$ among all subsampling-based methods, while even under $C=100$ its AUC remains comparable to competing methods. The performance gaps across methods become more substantial as the sampling budget decreases, suggesting that $\text{CE}$ provides greater improvement in resource-limited settings.


\subsection{Stroke prediction}\label{sec:data_example2}

Another experiment was designed to further assess the practical performance. In this section, we examine a scenario in which the prevalence is low as $\Pr(Y=1)\approx 5\%$ and no surrogate is available, to evaluate the effectiveness of the proposed method under rare response conditions and to determine whether it can operate robustly without surrogate information. In this application, we set $S=0$ for all observations, so that all records belong to the unlabeled observations and the proposed method reduces to a covariate-assisted optimal sampling design without surrogate-positive cases.

The dataset was obtained from Kaggle and has also been analyzed in recent studies \citep{chakraborty2024predicting, melnykova2025machine}. The objective of this analysis is to predict the risk of stroke based on patients' records. Accordingly, the binary response is defined as $Y=1$ for patients who have experienced a stroke and $Y=0$ otherwise. The dataset contains 5,110 patient records and 12 attributes, including demographic, clinical, and lifestyle factors relevant to stroke risk. Data preprocessing was conducted to remove observations containing unavailable or irrelevant information. Records with missing BMI values (``N/A'') were removed, reducing the dataset to 4,909 observations. One record labeled ``Other'' in the gender variable was removed. Observations with an ``Unknown'' smoking status were excluded because this label indicates missing smoking information. After preprocessing, a total of 3,425 complete observations remained for analysis. All methods were evaluated under sampling budgets $C=200, 300, 400$ with $m=300$ and 1,000 replicates.

\begin{table}[t]
\centering
\caption{Summary of real data application in Section \ref{sec:data_example2}}
\label{tab:combined_results_realdata2}
\begin{tabular}{llrrrrr}
\toprule
\textbf{C} & \textbf{Method} & \textbf{CE} & \textbf{MSE} & \textbf{TN} & \textbf{TP} & \textbf{AUC} \\
\midrule

\multirow{5}{*}{200}
& CE   & 731.343 & 0.054 & 0.985 & 0.079 & 0.753 \\
& MSE  & N/A     & 0.054 & 1.000 & 0.000 & 0.500 \\
& OSCA & 963.944 & 0.057 & 0.980 & 0.092 & 0.714 \\
& SRS  & 963.944 & 0.057 & 0.980 & 0.092 & 0.714 \\
& FULL & 531.262 & 0.046 & 1.000 & 0.006 & 0.833 \\
\midrule

\multirow{5}{*}{300}
& CE   & 627.556 & 0.051 & 0.991 & 0.055 & 0.782 \\
& MSE  & N/A     & 0.054 & 1.000 & 0.000 & 0.500 \\
& OSCA & 679.059 & 0.052 & 0.989 & 0.061 & 0.761 \\
& SRS  & 679.059 & 0.052 & 0.989 & 0.061 & 0.761 \\
& FULL & 531.262 & 0.046 & 1.000 & 0.006 & 0.833 \\
\midrule

\multirow{5}{*}{400}
& CE   & 590.704 & 0.050 & 0.994 & 0.042 & 0.796 \\
& MSE  & N/A     & 0.054 & 1.000 & 0.000 & 0.500 \\
& OSCA & 613.811 & 0.050 & 0.993 & 0.048 & 0.788 \\
& SRS  & 613.811 & 0.050 & 0.993 & 0.048 & 0.788 \\
& FULL & 531.262 & 0.046 & 1.000 & 0.006 & 0.833 \\
\bottomrule
\end{tabular}
\end{table}

In Table \ref{tab:combined_results_realdata2}, we summarize the performance of the proposed and competing methods under different sampling budgets ($C = 200, 300, 400$). Among the subsampling-based methods, the proposed CE method consistently achieves the lowest cross-entropy loss and the highest AUC, demonstrating stable prediction performance under rare-outcome conditions. OSCA exhibits exactly the same performance as SRS because no surrogate is available in this analysis, so $\Pr(S=1 \mid \Z)$ cannot be estimated and the OSCA calibration reduces to simple random sampling. In contrast, the MSE method produces numerically degenerate fitted probabilities, leading to invalid CE values and an AUC of 0.5. Overall, the proposed CE method remains effective even under low-prevalence conditions without surrogate information, achieving the performance closest to the FULL benchmark across the considered budgets.


\section{Discussion}\label{sec:discussion}

We proposed a sampling procedure for risk prediction under measurement constraints, optimizing the expected cross-entropy loss using data that may include a positive-only surrogate of the response. Our method is applicable to a broad class of smooth parametric prediction models and does not require responses from the target unlabeled observations when constructing the sampling probabilities, once a preliminary estimator is available. Theoretical results establish consistency, asymptotic normality, and leading-order cross-entropy optimality of the proposed procedure. Empirical studies, including simulations and two real-world applications, demonstrate the superior prediction performance of our method across scenarios, including empirical robustness checks under false-positive surrogates and extremely low response prevalence.

While the proposed procedure demonstrates strong theoretical and empirical properties, there are several practical considerations and avenues for future research. The invertibility of the expected Hessian matrix may become challenging in high-dimensional settings, particularly when the number of covariates is large or when strong collinearity is present. In such cases, identifying informative covariates or incorporating regularization techniques may help to ensure stable estimation. Additionally, although we treated the pilot sample as separate from the second-stage sample for theoretical clarity, in practice, it can be incorporated into the final training set, thereby improving data efficiency. A rigorous theoretical justification of this implementation would require dedicated investigation, and establishing such results remains an important direction for future work.


\section*{Supplementary material}

The supplementary material includes all of the technical details.


\section*{Data availability}

The data analyzed in Section~\ref{sec:realdata} consist of two datasets. The depression dataset is publicly available at the supplementary material of \cite{lee2023optimal}. The stroke dataset  is available at the following URL: https://www.kaggle.com/datasets/fedesoriano/stroke-prediction-dataset.


\section*{Acknowledgments}
This work was supported by the National Research Foundation of Korea(NRF) grant funded by the Korea government(MSIT) (No. RS-2026-25468474).


\bibliographystyle{agsm}
\bibliography{optsampleentropy}


\clearpage
\pagenumbering{arabic}
\section*{Supplementary material}
\setcounter{equation}{0}\renewcommand{\theequation}{S.\arabic{equation}}
\setcounter{subsection}{0}\renewcommand{\thesubsection}{S.\arabic{subsection}}

\subsection{Proof of Theorem 1}\label{sm:pf1}
$\wh\bb_\pi$ optimizes $(1)$ hence satisfies
\begin{align*}
    \0
    =&~-N^{-1}\frac{\partial\mathcal{L}_N(\wh\bb_\pi)}{\partial\bb}\\
    =&~N^{-1}\sumI\frac{\delta_i}{\pi_{\rm T}(\x_i)}\left\{\frac{y_i}{p(\x_i,\wh\bb_\pi)}-\frac{1-y_i}{1-p(\x_i,\wh\bb_\pi)}\right\}\p_{\bb}'(\x_i,\wh\bb_\pi),
\end{align*}
i.e., $\wh\bb_\pi$ is a solution to $N^{-1}\sumI \mathbf{g}(\delta_i,\delta_iy_i,\x_i,\bb)=\0$, where
\begin{align*}
    \mathbf{g}(\delta_i,\delta_iy_i,\x_i,\bb)=\frac{\delta_i}{\pi_{\rm T}(\x_i)}\left\{\frac{y_i}{p(\x_i,\bb)}-\frac{1-y_i}{1-p(\x_i,\bb)}\right\}\p_{\bb}'(\x_i,\bb).
\end{align*}
Now, to apply Theorem 2.1 of Newey $\&$ McFadden $(1994)$, let us define 
\begin{align*}
    \wh Q_N(\bb)=-\left\|N^{-1}\sumI \mathbf{g}(\delta_i,\delta_iy_i,\x_i,\bb)\right\|_2^2,\quad
    Q_0(\bb)=-\left\|\E\{\mathbf{g}(\Delta,\Delta Y,\X,\bb)\}\right\|_2^2.
\end{align*}
Then, (i) $Q_0(\bb)$ is uniquely maximized at $\bb=\bb_0$ since
\begin{align*}
    \E\{\mathbf{g}(\Delta,\Delta Y,\X,\bb_0)\}
    =&~\E\left[\frac{\Delta}{\pi_{\rm T}(\X)}\left\{\frac{Y}{p(\X,\bb_0)}-\frac{1-Y}{1-p(\X,\bb_0)}\right\}\p_{\bb}'(\X,\bb_0)\right]\\
    =&~\E\left[\frac{\pi_{\rm T}(\X)}{\pi_{\rm T}(\X)}\left\{\frac{p(\X,\bb_0)}{p(\X,\bb_0)}-\frac{1-p(\X,\bb_0)}{1-p(\X,\bb_0)}\right\}\p_{\bb}'(\X,\bb_0)\right]\\
    =&~\0,
\end{align*}
and
\begin{align*}
    \frac{\partial}{\partial\bb}\E\{\mathbf{g}(\Delta,\Delta Y,\X,\bb_0)\}
    =&~\E\left\{\frac{\Delta}{\pi_{\rm T}(\X)}\left(\left[-\frac{Y}{\{p(\X,\bb_0)\}^2}-\frac{1-Y}{\{1-p(\X,\bb_0)\}^2}\right]{\p_{\bb}'}^{\otimes2}(\X,\bb_0)\right.\right.\\
    &~\left.\left.+\left\{\frac{Y}{p(\X,\bb_0)}-\frac{1-Y}{1-p(\X,\bb_0)}\right\}\p_{\bb\bb\trans}''(\X,\bb_0)\right)\right\}\\
    =&~-\E\left[\left\{\frac{1}{p(\X,\bb_0)}+\frac{1}{1-p(\X,\bb_0)}\right\}{\p_{\bb}'}^{\otimes2}(\X,\bb_0)\right]\\
    =&~-\D
\end{align*}
is invertible by Condition $(\text{C}3)$. (ii) $\bb_0\in\Theta$ and $\Theta$ is compact by Condition $(\text{C}1)$. (iii) $Q_0(\bb)$ is continuous by Condition$(\text{C}2)$. (iv) $\wh Q_N(\bb)$ converges uniformly in probability to $Q_0(\bb)$ because
\begin{align*}
    N^{-1}\sumI \mathbf{g}(\delta_i,\delta_iy_i,\x_i,\bb)=N^{-1}\sumI \frac{\delta_i}{\pi_{\rm T}(\x_i)}\left\{\frac{y_i}{p(\x_i,\bb)}-\frac{1-y_i}{1-p(\x_i,\bb)}\right\}\p_{\bb}'(\x_i,\bb)
\end{align*}
converges in probability to $\E\{\mathbf{g}(\Delta,\Delta Y,\X,\bb)\}$ by the law of large numbers under Conditions $(\text{C}1)$,$(\text{C}2)$, $(\text{C}4)$, and this convergence is uniform with respect to $\bb$ by Lemma 2.8 of Newey $\&$ McFadden $(1994)$ because $\bb\in\Theta$ and $\Theta$ is compact by Condition $(\text{C}1)$ and $\mathbf{g}(\delta,\delta y,\x,\bb)$ is equicontinuous with respect to $\bb$ by Condition$(\text{C}2)$. Therefore, by Theorem 2.1 of Newey $\&$ McFadden $(1994)$, $\wh\bb_\pi$ converges in probability to $\bb_0$.
\hfill$\blacksquare$

\subsection{Proof of Theorem 2}\label{sm:pf2}

$\wh\bb_\pi$ minimizes (1) subject to $\bb \in \Theta$, 
\begin{align*}
    \mathcal{L}_N(\bb)=-\sumI\frac{\delta_i}{\pi_{\rm T}(\x_i)}\left[y_i\log p(\x_i,\bb)+(1-y_i)\log\{1-p(\x_i,\bb)\}\right].
\end{align*}
To apply Theorem 3.1 of Newey $\&$ McFadden $(1994)$, let us define 
\begin{align*}
    \wh Q_N(\bb)
    =&~-N^{-1}\mathcal{L}_N(\bb)\\
    =&~N^{-1}\sumI\frac{\delta_i}{\pi_{\rm T}(\x_i)}\left[y_i\log p(\x_i,\bb)+(1-y_i)\log\{1-p(\x_i,\bb)\}\right].
\end{align*} 
Then, $\wh\bb_\pi$ equivalently maximizes $\wh Q_N(\bb)$ subject to $\bb \in \Theta$. 
By Theorem 1, $\wh\bb_\pi \xrightarrow{p} \bb_0$. (i) $\bb_0 \in \operatorname{int}(\Theta)$ by Condition $(\text{C}1)$. (ii) $\wh Q_N(\bb)$ is twice continuously differentiable in a neighborhood $\mathcal{N}$ of $\bb_0$ by Condition$(\text{C}2)$. (iii) $\sqrt{N}\nabla_{\bb}\wh Q_N(\bb_0) \xrightarrow{d} N(\0,\bSig_\pi)$ because
\begin{align*}
    \nabla_{\bb}\wh Q_N(\bb_0)
    =&~N^{-1}\sumI\frac{\delta_i}{\pi_{\rm T}(\x_i)}\left\{\frac{y_i}{p(\x_i,\bb_0)}-\frac{1-y_i}{1-p(\x_i,\bb_0)}\right\}\p_{\bb}'(\x_i,\bb_0)\\
    =&~N^{-1}\sumI\mathbf{g}(\delta_i,\delta_iy_i,\x_i,\bb_0),
\end{align*}
by the law of iterated expectations, we obtain
\begin{align*}
    \E\{\mathbf{g}(\Delta,\Delta Y,\X,\bb_0)\}
    =&~\E\left[\frac{\Delta}{\pi_{\rm T}(\X)}\left\{\frac{Y}{p(\X,\bb_0)}-\frac{1-Y}{1-p(\X,\bb_0)}\right\}\p_{\bb}'(\X,\bb_0)\right]\\
    =&~\E\left[\frac{\pi_{\rm T}(\X)}{\pi_{\rm T}(\X)}\left\{\frac{p(\X,\bb_0)}{p(\X,\bb_0)}-\frac{1-p(\X,\bb_0)}{1-p(\X,\bb_0)}\right\}\p_{\bb}'(\X,\bb_0)\right]\\
    =&~\0,
\end{align*}
where the second equality follows from $\E(\Delta \mid Y,\X)=\E(\Delta \mid \X)=\pi_{\rm T}(\X)$, $\E[Y - p(\mathbf{X}, \bb_0) \mid \mathbf{X}]=0$, and Condition $(\text{C}1)$.
In addition, 
\begin{align*}
    \V\left[\mathbf{g}(\Delta,\Delta Y,\X,\bb_0)\right]
    =&~\V\left[\frac{\Delta}{\pi_{\rm T}(\X)}\left\{\frac{Y}{p(\X,\bb_0)}-\frac{1-Y}{1-p(\X,\bb_0)}\right\}\p_{\bb}'(\X,\bb_0)\right]\\
    =&~\E\left[\frac{\Delta^2}{\{\pi_{\rm T}(\X)\}^2}\left\{\frac{Y}{p(\X,\bb_0)}-\frac{1-Y}{1-p(\X,\bb_0)}\right\}^2{\p_{\bb}'}^{\otimes2}(\X,\bb_0)\right]\\
    =&~\E\left[\frac{\Delta}{\{\pi_{\rm T}(\X)\}^2}\left[\frac{Y}{\{p(\X,\bb_0)\}^2}+\frac{1-Y}{\{1-p(\X,\bb_0)\}^2}\right]{\p_{\bb}'}^{\otimes2}(\X,\bb_0)\right]\\
    =&~\E\left[\frac{1}{\pi_{\rm T}(\X)}\left\{\frac{1}{p(\X,\bb_0)}+\frac{1}{1-p(\X,\bb_0)}\right\}{\p_{\bb}'}^{\otimes2}(\X,\bb_0)\right]\\
    =&~\E\left[\frac{1}{\pi_{\rm T}(\X)}\frac{{\p_{\bb}'}^{\otimes2}(\X,\bb_0)}{p(\X,\bb_0)\{1-p(\X,\bb_0)\}}\right]\\
    =\bSig_\pi,
\end{align*}
and the $\ell_2$ norm of $\bSig_\pi$ is finite by Conditions $(\text{C}1)$,$(\text{C}2)$, and $(\text{C}4)$. The third equality follows from $Y\in\{0,1\}$ and $\Delta\in\{0,1\}$, and the fourth equality follows from $\E(\Delta \mid Y,\X)=\E(\Delta \mid \X)=\pi_{\rm T}(\X)$, $\E[Y - p(\mathbf{X}, \bb_0) \mid \mathbf{X}]=0$, and Condition $(\text{C}1)$.
Therefore, the asymptotic distribution follows by the multivariate central limit theorem. 
(iv) There is $\H(\bb)$ that is continuous at $\bb_0$ and $\sup_{\bb\in\mathcal{N}}\|\nabla_{\bb\bb\trans} \wh Q_N(\bb)-\H(\bb)\|_2\xrightarrow{p} 0$. To show this, we apply Lemma 2.4 of Newey $\&$ McFadden $(1994)$.
We can express $\nabla_{\bb\bb\trans}\wh Q_N(\bb)$ as 
\begin{align*}
    \nabla_{\bb\bb\trans}\wh Q_N(\bb)=N^{-1}\sumI \a(\delta_i,\delta_i y_i,\x_{i},\bb)
\end{align*}
and define $\H(\bb)=\E[\a(\Delta,\Delta Y,\X,\bb)]$, where
\begin{align*}
    \a(\delta_i,\delta_i y_i,\x_{i},\bb)
    =&\frac{\delta_i}{\pi_{\rm T}(\x_i)}\left(\left[-\frac{y_i}{\{p(\x_i,\bb)\}^2}-\frac{1-y_i}{\{1-p(\x_i,\bb)\}^2}\right]{\p_{\bb}'}^{\otimes2}(\x_i,\bb)\right.\\
    &\left.+\left\{\frac{y_i}{p(\x_i,\bb)}-\frac{1-y_i}{1-p(\x_i,\bb)}\right\}\p_{\bb\bb\trans}''(\x_i,\bb)\right).
\end{align*}

By Conditions $(\text{C}1)$ and$(\text{C}2)$, $\a(\delta_i,\delta_i y_i,\x_{i},\bb)$ is continuous in $\bb$ on the compact set $\Theta$.
Moreover, $\|\a(\delta_i,\delta_i y_i,\x_i,\bb)\|_2$ admits a dominating function $d(\x)$ such that
$\|\a(\delta,\delta y,\x,\bb)\|_2\le d(\x)$ for all $\bb\in\Theta$ and $\E[d(\x)]<\infty$ by Conditions $(\text{C}1)$,$(\text{C}2)$, and $(\text{C}4)$, where
\begin{align*}
    d(\x_i)=\frac{1}{\epsilon_\pi}\Bigg(\frac{1}{{\epsilon^2_p}}\sup_{\bb\in\Theta}\|{\p_{\bb}'}(\x_i,\bb)\|_2^2+\frac{1}{\epsilon_p}\sup_{\bb\in\Theta}\|\p_{\bb\bb\trans}''(\x_i,\bb)\|_2\Bigg).
\end{align*} 
Indeed,
\begin{align*}
    \|\a(\delta_i,\delta_i y_i,\x_{i},\bb)\|_2
    =&~ \Bigg\|\frac{\delta_i}{\pi_{\rm T}(\x_i)}\left(\left[-\frac{y_i}{\{p(\x_i,\bb)\}^2}-\frac{1-y_i}{\{1-p(\x_i,\bb)\}^2}\right]{\p_{\bb}'}^{\otimes2}(\x_i,\bb)\right.\\
    &\left.+\left\{\frac{y_i}{p(\x_i,\bb)}-\frac{1-y_i}{1-p(\x_i,\bb)}\right\}\p_{\bb\bb\trans}''(\x_i,\bb)\right)\Bigg\|_2\\
    \le&~  \frac{1}{\epsilon_\pi}\left(\Bigg\|\left[-\frac{y_i}{\{p(\x_i,\bb)\}^2}-\frac{1-y_i}{\{1-p(\x_i,\bb)\}^2}\right]{\p_{\bb}'}^{\otimes2}(\x_i,\bb)\Bigg\|_2\right.\\
    &\left. +\Bigg\|\left\{\frac{y_i}{p(\x_i,\bb)}-\frac{1-y_i}{1-p(\x_i,\bb)}\right\}\p_{\bb\bb\trans}''(\x_i,\bb)\Bigg\|_2\right)\\
    \le&~  \frac{1}{\epsilon_\pi}\Bigg(\frac{1}{{\epsilon^2_p}}\|{\p_{\bb}'}(\x_i,\bb)\|_2^2+\frac{1}{\epsilon_p}\|\p_{\bb\bb\trans}''(\x_i,\bb)\|_2\Bigg).
\end{align*}
The first inequality is obtained by applying the triangle inequality, using the fact that $\delta_i\in\{0,1\}$, and Condition $(\text{C}4)$. The second inequality follows from $y_i\in\{0,1\}$ and Condition $(\text{C}1)$. Note also that $\|{\p_{\bb}'}^{\otimes2}(\x_i,\bb)\|_2=\|{\p_{\bb}'}(\x_i,\bb)\|_2^2$ under the spectral norm. 
Lemma 2.4 of Newey $\&$ McFadden $(1994)$ then implies that $\E\left[\a(\Delta,\Delta Y,\X,\bb)\right]$ is continuous and $$\sup_{\bb\in\Theta}\left\|N^{-1}\sumI \a(\delta_i,\delta_i y_i,\x_{i},\bb)-\E\left[\a(\Delta,\Delta Y,\X,\bb)\right]\right\|_2\xrightarrow{p} 0.$$
(v) $\H=\H(\bb_0)$ is nonsingular because 
\begin{align*}
    \H(\bb_0)
    =&~\E\left\{\frac{\Delta}{\pi_{\rm T}(\X)}\left(\left[-\frac{Y}{\{p(\X,\bb_0)\}^2}-\frac{1-Y}{\{1-p(\X,\bb_0)\}^2}\right]{\p_{\bb}'}^{\otimes2}(\X,\bb_0)\right.\right.\\
    &\left.\left.+\left\{\frac{Y}{p(\X,\bb_0)}-\frac{1-Y}{1-p(\X,\bb_0)}\right\}\p_{\bb\bb\trans}''(\X,\bb_0)\right)\right\}\\
    =&~-\E\left[\left\{\frac{1}{p(\X,\bb_0)}+\frac{1}{1-p(\X,\bb_0)}\right\}{\p_{\bb}'}^{\otimes2}(\X,\bb_0)\right]\\
    =&~-\D
\end{align*}
is invertible by Condition $(\text{C}3)$, where the second equality follows from $\E(\Delta \mid Y,\X)=\E(\Delta \mid \X)=\pi_{\rm T}(\X)$ and $\E[Y - p(\mathbf{X}, \bb_0) \mid \mathbf{X}]=0$. As a result, by Theorem 3.1 of Newey $\&$ McFadden $(1994)$, $$\sqrt{N}(\wh\bb_\pi - \bb_0) \xrightarrow{d} N(\0,\D^{-1}\bSig_\pi \D^{-1}),$$ where
\begin{align*}
    \D
    =\E\left[\frac{{\p_{\bb}'}^{\otimes2}(\X,\bb_0)}{p(\X,\bb_0)\{1-p(\X,\bb_0)\}}\right],\quad
    \bSig_\pi
    =\E\left[\frac{1}{\pi_{\rm T}(\X)}\frac{{\p_{\bb}'}^{\otimes2}(\X,\bb_0)}{p(\X,\bb_0)\{1-p(\X,\bb_0)\}}\right].
\end{align*}
\hfill$\blacksquare$

\subsection{Proof of Theorem 3} \label{sm:pf3}

By the Taylor expansion of $Q(\bb)$ around $\bb_0$ under Conditions $(\text{C}1)$ and$(\text{C}2)$, we can write the expected cross-entropy loss as
\begin{align}\label{eq:loss1}
    Q(\wh\bb_\pi)
    =&~-\E\left[Y^*\log p(\X^*,\wh\bb_\pi)+(1-Y^*)\log\{1-p(\X^*,\wh\bb_\pi)\}\right]\n\\
    =&~-\E\Big(Y^*\log p(\X^*,\bb_0)+(1-Y^*)\log\{1-p(\X^*,\bb_0)\}\n\\
    &+\left\{\frac{Y^*}{p(\X^*,\bb_0)}-\frac{1-Y^*}{1-p(\X^*,\bb_0)}\right\}{\p_{\bb}'}\trans(\X^*,\bb_0)(\wh\bb_\pi-\bb_0)\n\\
    &+\frac{1}{2}\left[-\frac{Y^*}{\{p(\X^*,\bb_0)\}^2}-\frac{1-Y^*}{\{1-p(\X^*,\bb_0)\}^2}\right](\wh\bb_\pi-\bb_0)\trans{\p_{\bb}'}^{\otimes2}(\X^*,\bb_0)(\wh\bb_\pi-\bb_0)\n\\
    &+\frac{1}{2}\left\{\frac{Y^*}{p(\X^*,\bb_0)}-\frac{1-Y^*}{1-p(\X^*,\bb_0)}\right\}(\wh\bb_\pi-\bb_0)\trans\p''_{\bb\bb\trans}(\X^*,\bb_0)(\wh\bb_\pi-\bb_0)\Big)\n\\
    &+O\{\E(\|\wh\bb_\pi-\bb_0\|_2^3)\}\n\\
    =&~-\E\left[Y^*\log p(\X^*,\bb_0)+(1-Y^*)\log\{1-p(\X^*,\bb_0)\}\right]\n\\
    &+\frac{1}{2}\E\left[\left\{\frac{1}{p(\X^*,\bb_0)}+\frac{1}{1-p(\X^*,\bb_0)}\right\}(\wh\bb_\pi-\bb_0)\trans{\p_{\bb}'}^{\otimes2}(\X^*,\bb_0)(\wh\bb_\pi-\bb_0)\right]\n\\
    &+O\{\E(\|\wh\bb_\pi-\bb_0\|_2^3)\},
\end{align}
where the third equality follows since $\X^*$ is independent of $\wh\bb_\pi$ and $\E(Y^*\mid \X^*)= p(\mathbf{X}^*, \bb_0)$. Here, the second term can be expressed as 
\begin{align}\label{eq:loss2}
    &~\E\left[\left\{\frac{1}{p(\X^*,\bb_0)}+\frac{1}{1-p(\X^*,\bb_0)}\right\}(\wh\bb_\pi-\bb_0)\trans{\p_{\bb}'}^{\otimes2}(\X^*,\bb_0)(\wh\bb_\pi-\bb_0)\right]\n\\
    =&~\E\left[\frac{(\wh\bb_\pi-\bb_0)\trans{\p_{\bb}'}^{\otimes2}(\X^*,\bb_0)(\wh\bb_\pi-\bb_0)}{p(\X^*,\bb_0)\{1-p(\X^*,\bb_0)\}}\right]\n\\
    =&~\tr\left(\E\left[\frac{{\p_{\bb}'}^{\otimes2}(\X^*,\bb_0)}{p(\X^*,\bb_0)\{1-p(\X^*,\bb_0)\}}(\wh\bb_\pi-\bb_0)^{\otimes2}\right]\right)\n\\
    =&~\tr\left(\E\left[\frac{{\p_{\bb}'}^{\otimes2}(\X^*,\bb_0)}{p(\X^*,\bb_0)\{1-p(\X^*,\bb_0)\}}\right]\E\left\{(\wh\bb_\pi-\bb_0)^{\otimes2}\right\}\right)\n\\
    =&~\tr\left(\D\E\left\{(\wh\bb_\pi-\bb_0)^{\otimes2}\right\}\right),
\end{align}
where the second equality holds because $\X^*$ is independent of $\wh\bb_\pi$.
Therefore, $Q(\wh\bb_\pi)$ can be written as
\begin{align*}
    Q(\wh\bb_\pi)
    =&~Q(\bb_0)+\frac{1}{2}\tr\left(\D\E\left\{(\wh\bb_\pi-\bb_0)^{\otimes2}\right\}\right)+O\{\E(\|\wh\bb_\pi-\bb_0\|_2^3)\}.
\end{align*}
Incorporating the asymptotic variance from Theorem 2, we get
\begin{align*}
    N\{ Q(\wh\bb_\pi) - Q(\bb_0) \}\xrightarrow{p}\frac{1}{2}\tr(\bSig_\pi\D^{-1})
\end{align*}
as $N \to \infty$. Next, we express $\tr(\bSig_\pi\D^{-1})$ as
\begin{align}\label{eq:loss3}
    \tr(\bSig_\pi\D^{-1})
    =&~\tr\left(\E\left[\frac{1}{\pi_{\rm T}(\X)}\frac{{\p_{\bb}'}^{\otimes2}(\X,\bb_0)}{p(\X,\bb_0)\{1-p(\X,\bb_0)\}}\right]\D^{-1}\right)\n\\
    =&~\E\left[\frac{1}{S+(1-S)\pi(\Z)}\frac{{\p_{\bb}'}\trans(\X,\bb_0)\D^{-1}{\p_{\bb}'}(\X,\bb_0)}{p(\X,\bb_0)\{1-p(\X,\bb_0)\}}\right]\n\\
    =&~\E\left[\frac{{\p_{\bb}'}\trans(\X,\bb_0)\D^{-1}{\p_{\bb}'}(\X,\bb_0)}{p(\X,\bb_0)\{1-p(\X,\bb_0)\}}\mid S=1\right]\Pr(S=1)\n\\
    &+\E\left[\frac{1}{\pi(\Z)}\frac{{\p_{\bb}'}\trans(\X,\bb_0)\D^{-1}{\p_{\bb}'}(\X,\bb_0)}{p(\X,\bb_0)\{1-p(\X,\bb_0)\}}\mid S=0\right]\Pr(S=0).
\end{align}
Here, the first term does not involve $\pi(\cdot)$, and the empirical version of the second term can be expressed as 
\begin{align}\label{eq:loss4}
    &~N^{-1}\sum_{i:s_i=0}\frac{1}{\pi(\z_i)}\frac{{\p_{\bb}'}\trans(\x_i,\wt\bb)\wt\D^{-1}{\p_{\bb}'}(\x_i,\wt\bb)}{p(\x_i,\wt\bb)\{1-p(\x_i,\wt\bb)\}}\\
    =&~\tr\Bigg(\wt\D^{-1}\Bigg[N^{-1}\sumI \mathbb{I}(s_i=0)\frac{1}{\pi(\z_i)}\frac{{\p_{\bb}'}^{\otimes 2}(\x_i,\wt\bb)}{p(\x_i,\wt\bb)\{1-p(\x_i,\wt\bb)\}}\Bigg]\Bigg)\n\\
    =&~\tr\left(\wt\D^{-1}\wt\A\right),\n
\end{align}
where $\wt\bb$ is a consistent estimate of $\bb_0$,
\begin{align*}
\wt\D = N^{-1}\sumI \frac{{\p_{\bb}'}^{\otimes2}(\x_i,\wt\bb)}{p(\x_i,\wt\bb)\{1-p(\x_i,\wt\bb)\}}, \quad
\wt\A = N^{-1}\sum_{i=1}^N \mathbb{I}(s_i=0) \frac{1}{\pi(\z_i)}\frac{{\p_{\bb}'}^{\otimes 2}(\x_i,\wt\bb)}{p(\x_i,\wt\bb)\{1-p(\x_i,\wt\bb)\}}.
\end{align*}
To show the convergence of both $\wt\D$ and $\wt\A$, we apply Lemma 4.3 of Newey $\&$ McFadden $(1994)$. First, let us define
\begin{align*}
    \a_\D(\x,\bb)
    =\frac{{\p_{\bb}'}^{\otimes2}(\x,\bb)}{p(\x,\bb)\{1-p(\x,\bb)\}}.
\end{align*}
By Conditions $(\text{C}1)$ and$(\text{C}2)$, $\a_\D(\x,\bb)$ is continuous at $\bb_0$. Moreover, there is a neighborhood $\mathcal{N}$ of $\bb_0$ such that $\E[\sup_{\bb \in \mathcal{N}}\|\a_\D(\X,\bb)\|_2]< \infty$ by Conditions $(\text{C}1)$ and$(\text{C}2)$.
Indeed, by applying the properties of the outer product norm and using Condition $(\text{C}1)$,
\begin{align*}
    \|\a_\D(\x,\bb)\|_2
    &~=\Bigg\|\frac{{\p_{\bb}'}^{\otimes2}(\x,\bb)}{p(\x,\bb)\{1-p(\x,\bb)\}}\Bigg\|_2\\
    &~ \le \frac{1}{\epsilon_p (1-\epsilon_p)}\|{\p_{\bb}'}(\x,\bb)\|_2^2.
\end{align*}
Since $\sup_{\bb \in \mathcal{N}}\|{\p_{\bb}'}(\x,\bb)\|_2^2$ is bounded under Condition$(\text{C}2)$, the dominance condition is satisfied. Therefore, by Lemma 4.3 of Newey $\&$ McFadden $(1994)$, we obtain $\wt\D \xrightarrow{p} \D$. Similarly, to show the convergence of $\wt\A$, we define
\begin{align*}
    \a_\A(\x,\bb) = \frac{1}{\pi(\z)}\frac{{\p_{\bb}'}^{\otimes 2}(\x,\bb)}{p(\x,\bb)\{1-p(\x,\bb)\}}.
\end{align*}
By Conditions $(\text{C}1)$ and$(\text{C}2)$, $\a_\A(\x,\bb)$ is continuous at $\bb_0$. Furthermore, there is a neighborhood $\mathcal{N}$ of $\bb_0$ such that $\E[\sup_{\bb \in \mathcal{N}}\|\a_\A(\X,\bb)\|_2]< \infty$ by Conditions $(\text{C}1)$,$(\text{C}2)$, and $(\text{C}4)$. Indeed, by applying the properties of the outer product norm and using Condition $(\text{C}1)$ and $(\text{C}4)$, its norm is bounded by
\begin{align*}
    \|\a_\A(\x,\bb)\|_2 
    &~=\Bigg\|\frac{1}{\pi(\z)}\frac{{\p_{\bb}'}^{\otimes 2}(\x,\bb)}{p(\x,\bb)\{1-p(\x,\bb)\}}\Bigg\|_2\\
    &~ \le\frac{1}{\epsilon_\pi}\frac{1}{\epsilon_p (1-\epsilon_p)}\|{\p_{\bb}'}(\x,\bb)\|_2^2.
\end{align*}
Since $\sup_{\bb \in \mathcal{N}}\|{\p_{\bb}'}(\x,\bb)\|_2^2$ is bounded under Condition$(\text{C}2)$, the dominance condition is satisfied. Therefore, by Lemma 4.3 of Newey $\&$ McFadden $(1994)$, we obtain $\wt\A \xrightarrow{p} \A$. Finally, by the continuous mapping theorem and Slutsky's theorem under Condition $(\text{C}3)$, we obtain
\begin{align*}
    \tr\left(\wt\D^{-1}\wt\A\right) \xrightarrow{p}\tr\left(\D^{-1}\A\right).
\end{align*}
Combining $(\text{S}.1)$, $(\text{S}.2)$, $(\text{S}.3)$ and $(\text{S}.4)$,
\begin{align*}
    &~-\E\left[Y^*\log p(\X^*,\wh\bb_\pi)+(1-Y^*)\log\{1-p(\X^*,\wh\bb_\pi)\}\right]\\
    =&~\frac{1}{2}N^{-2}\sum_{i:s_i=0}\frac{1}{\pi(\z_i)}\frac{{\p_{\bb}'}\trans(\x_i,\wt\bb)\wt\D^{-1}{\p_{\bb}'}(\x_i,\wt\bb)}{p(\x_i,\wt\bb)\{1-p(\x_i,\wt\bb)\}}+r+o_p(N^{-1}),
\end{align*}
where $r$ denotes the remainder that does not depend on $\pi(\cdot)$.
Now, since
\begin{align*}
    \wh{\pi}
    ={\arg\min}_{\{\pi_i: s_i=0\}}
    \sum_{i:s_i=0}\frac{a_i(\wt\bb)}{\pi_i}
    \quad\text{subject to}\quad
    \sum_{i:s_i=0}\pi_i\leq C,
    \quad
    \epsilon_\pi\leq \pi_i\leq 1,
\end{align*}
the leading empirical term under $\wh\pi$ is no larger than that under any admissible $\pi$. Therefore,
\begin{align*}
    Q(\wh\bb_{\wh{\pi}})
    \leq Q(\wh\bb_{\pi})+o_p(N^{-1})
\end{align*}
for arbitrary admissible $\pi$ subject to $\sum_{i:s_i=0}\pi_i\leq C$.

\hfill$\blacksquare$


\end{document}